\documentclass{elsart}
\usepackage{amssymb,epsfig,graphicx}
\begin{document}
\begin{frontmatter}

\title{Disorder Induced Negative Magnetization in LaSrCoRuO$_{6 }$}

\author[gu]{P. S. R. Murthy}, \author[gu]{K. R. Priolkar\corauthref{krp}}\ead{krp@unigoa.ac.in},
\author[tifr]{P. A. Bhobe}, \author[barc]{A. Das}, \author[gu]{P. R. Sarode} and \author[tifr]{A. K. Nigam}
\corauth[krp]{Corresponding author}
\address[gu]{Department of Physics, Goa University, Goa, 403 206 India.}
\address[tifr]{Tata Institute of Fundamental Research, Homi Bhabha Road, Mumbai, 400 005 India}
\address[barc]{Solid State Physics Division, Bhabha Atomic Research Centre, Trombay, Mumbai 400 085 India}

\begin{abstract}
This paper reports effect of thermally induced disorder on the magnetic properties of LaSrCoRuO$_6$ double
perovskite. While the ordered sample is antiferromagnetic, the disordered sample exhibits negative values of
magnetization measured in low applied fields. Isothermal magnetization on this sample shows hysteresis due to
presence of ferromagnetic interactions. Based on neutron diffraction and X-ray Absorption Fine Structure (XAFS)
studies, these results have been interpreted to be due disorder in site occupancy of Co and Ru leading to
octahedral distortions and formation of Ru-O-Ru ferromagnetic linkages. Below 150K these ferromagnetic Ru spins
polarize the Co spins in a direction opposite to that of the applied field resulting in observed negative
magnetization.
\end{abstract}

\end{frontmatter}

\section{Introduction}
Ordering of  B-site cations in the double perovskites is known to play an important role in deciding the
magnetic, transport and structural properties of these systems. Sr$_{2}$FeMoO$_{6}$ and Sr$_{2}$FeReO$_{6}$ both
of which display large low field magnetoresistance are two good examples \cite{kob,gop}. The itinerancy and
ferrimagnetism in the above materials arise from a double exchange type of mechanism in which the ordering and
electronic configurations play a critical role \cite{dd}. The characteristics of this type of ordering stems from
the fact that it combines features of both ferromagnetic (FM) and antiferromagnetic (AF) systems. Moreover, if
more than two spin sub-lattices are involved, a new phenomena like temperature induced magnetization reversal can
emerge \cite{tas}. So far, apart from ferrimagnets \cite{bel} only few other families of oxides which include
layered ruthenates and manganites have exhibited temperature induced magnetization reversal
\cite{tas,ren,ren1,yosh,yosh1,ravi,ravi1,cao}.

The layered compounds, often characterized by a strong competition between antiferromagnetic and ferromagnetic
coupling and a complex interplay of spin, charge, and orbital degrees of freedom, are extremely sensitive to
small perturbations such as slight structural alterations. Sr$_2$YRuO$_6$ is a typical example. Here the negative
magnetization observed in low fields has been ascribed to two oppositely ordered ferromagnetic superexchange
interactions viz, Ru-O-O-Ru and Ru-O-Y-O-Ru \cite{ravi}. LaSrCoRuO$_{6}$ is a type of layered compound wherein
degree of B-site (Co and Ru) order can lead to intriguing magnetic and transport properties \cite{bos, mam}.
However, here unlike Y, Co is a magnetic ion and that has led to different interpretations of the nature of
magnetic order in this compound. LaSrCoRuO$_6$ was reported to be a 3D variable range hopping semiconductor with
magnetic ordering temperature of 157K \cite{kim}. Recent studies suggest that the compound is an antiferromagnet
with T$_{N}$ = 87K \cite{bos} or a spin glass \cite{mam} and the transition at 157K could be due to SrRuO$_3$
impurity. Apart from SrRuO$_3$ impurity phase, the sharp rise in magnetization can also be due to ferromagnetic
Ru-O-Ru interactions arising from antiphasic grain boundaries \cite{skim}. The studies conducted on LaSrCoRuO$_6$
so far by varying the composition ratio of A site ions (La and Sr) highlight the importance of different magnetic
interactions between Co$^{2+/3+}$ and Ru$^{4+/5+}$ in governing the magnetic ground state \cite{mam,tomes}. This
paper reports the effect of thermally induced site-occupancy disorder in LaSrCoRuO$_6$ on its magnetic
properties. The most notable feature here is the observation of negative magnetization at low applied fields in
the more disordered sample. The results have been explained on the basis of EXAFS data recorded at Co and Ru
K-edge to be due to presence of additional ferromagnetic interactions resulting from B-site disorder in an
otherwise antiferromagnetic lattice.

\section{Experimental}
Two polycrystalline samples of LaSrCoRuO$_{6}$ were synthesized by solid state reaction method by taking
stoichiometric amounts of La$_{2}$O$_{3}$, SrCO$_{3}$, Co(NO$_{3}$)$_{2}$ and RuO$_{2}$. These starting powders
were ground thoroughly, pressed into pellets and heated for a total of 48 hrs, one at 1200$^\circ$C and the other
at 1300$^\circ$C with three intermediate regrinding steps. The sample annealed at 1200$^\circ$ was quenched to
room temperature while the other was furnace cooled. Both the samples were deemed to be phase pure, as X-ray
diffraction (XRD) data collected on a Rigaku X-ray diffractometer in the range of $ 18^\circ \le 2\theta \le
80^\circ$ using CuK$\alpha$ radiation showed no impurity reflections. The diffraction patterns were Rietveld
refined using FULLPROF suite and structural parameters were obtained. Scanning Electron Microscopy energy
dispersive spectroscopy (SEM EDS) and Iodometric titrations were carried out on these samples to confirm the
cation and oxygen stoichiometry. The cation stoichiometries were close the expected value of 10 at.\% in case of
both the samples. The oxygen stoichiometry in case of 1200$^\circ$C and 1300$^\circ$C annealed samples were found
to be 5.96$\pm$0.02 and 5.99$\pm$0.01 respectively. DC magnetization was measured, both, as a function of
temperature and magnetic field using the Quantum Design SQUID magnetometer (MPMS-5S). M(T) was measured in an
applied field of 50 Oe and 1000 Oe in the temperature range of 5 to 300 K. The sample was initially cooled from
300K to 5 K in zero applied field and the data was recorded while warming up to 300 K in the applied magnetic
field (referred to as ZFC curve) and subsequent cooling (referred to as FC curve) back to 5 K. Magnetization as a
function of field was measured under sweep magnetic fields up to $\pm 5 T$ at various temperatures. Before each
M(H) was recorded, the sample was warmed to 300 K and cooled back to the desired temperature. Neutron diffraction
(ND) measurements were performed at room temperature (RT) and 20K and a wavelength of 1.24\AA~ using powder
diffractometer at Dhruva, Trombay. XAFS experiments at the Co and Ru K edge were performed in transmission mode
at room temperature using the beamline 12C at Photon Factory, Tsukuba, Japan.

\section{Results and Discussion}

\begin{figure}[c]
\centering
\includegraphics[scale=0.5]{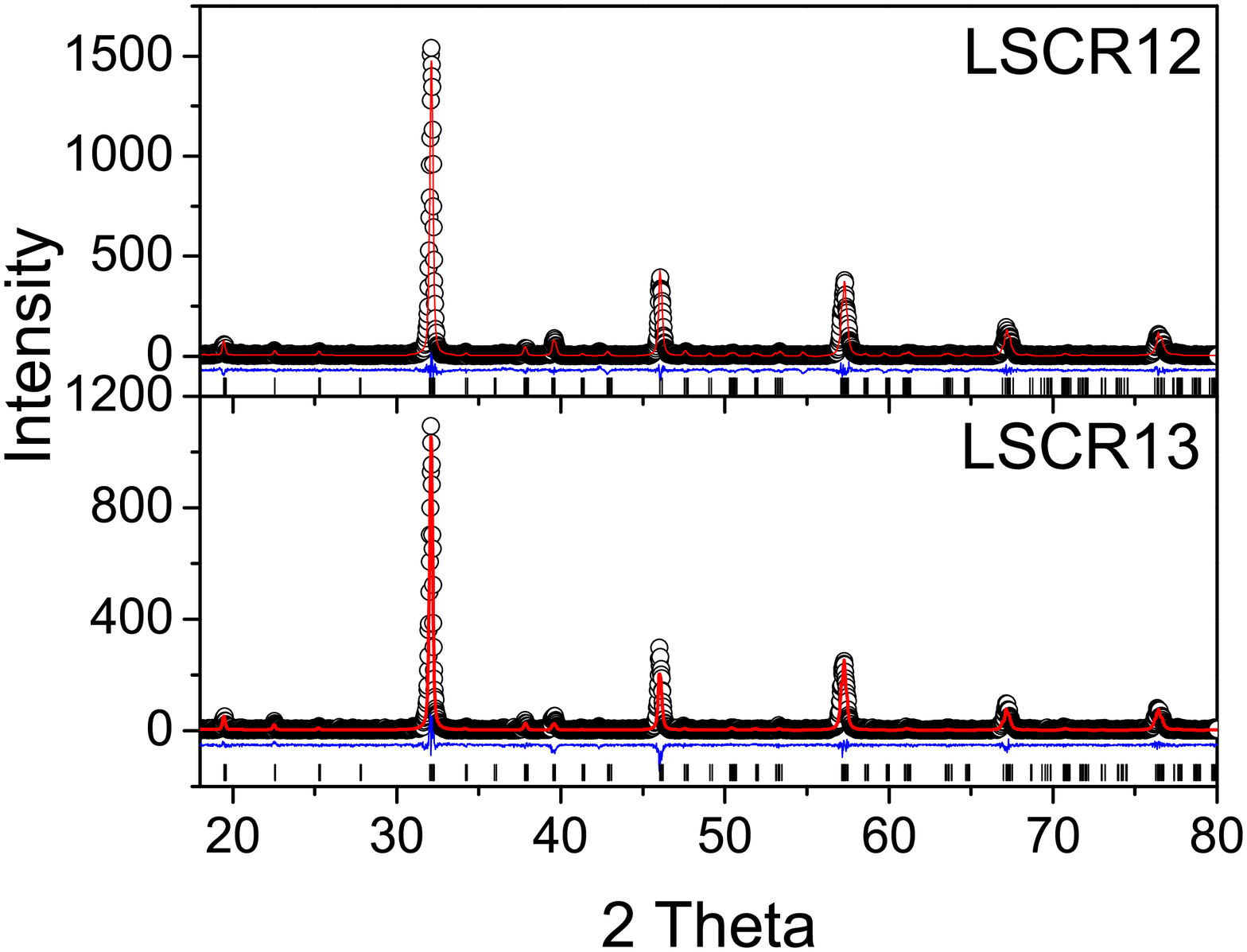}
\caption{\label{xrd06} Rietveld refined XRD patterns for LSCR13 and LSCR12. The open circles show the observed
counts and the continuous line passing through these counts is the calculated profile. The difference between the
observed and calculated patterns is shown as a continuous line at the bottom of the two profiles. The calculated
positions of the reflections are shown as vertical bars.}
\end{figure}

\begin{figure}[c]
\centering
\includegraphics[scale=0.5]{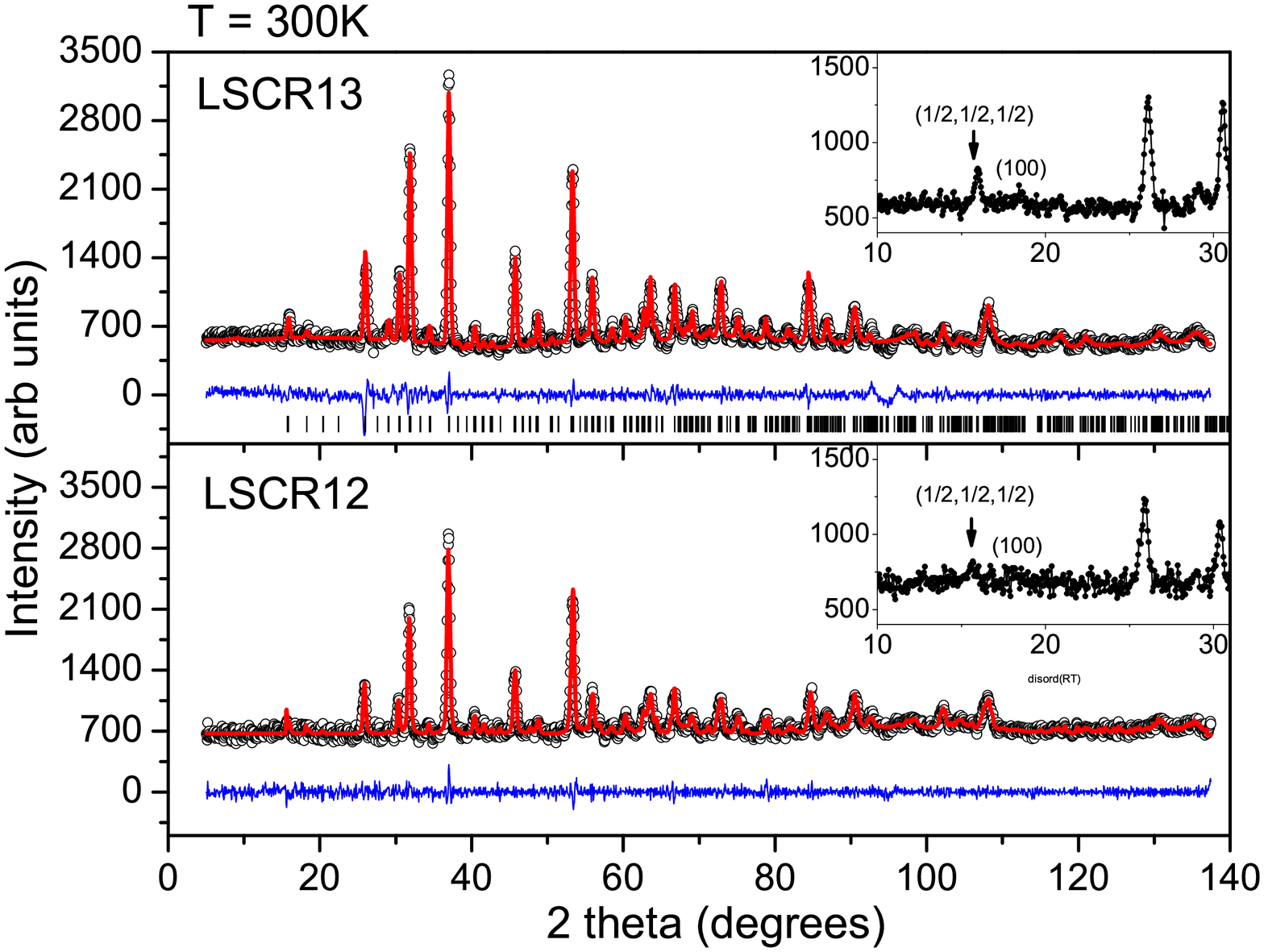}
\includegraphics[scale=0.5]{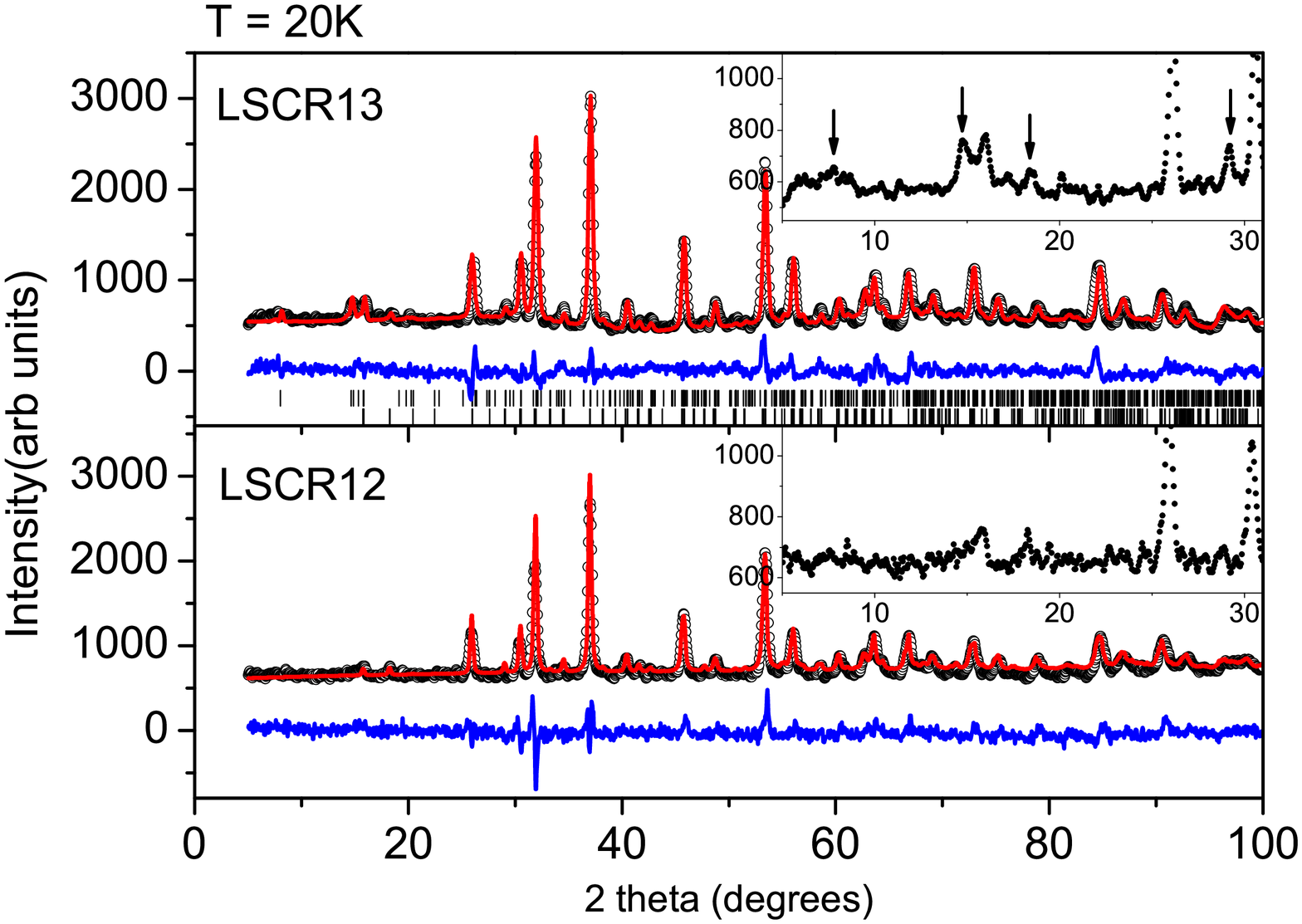}
\caption{\label{nd} Observed (circles), calculated (line) and difference ND patterns recorded at 300K (upper
panel) for LSCR13 and LSCR12. The inset presents data in limited range with the superlattice reflections seen
clearly in LSCR13 indicating higher degree of order. The lower panel shows neutron data taken at 20K for the same
samples. The inset presents data in limited range with the arrows indicating magnetic reflections present in
LSCR13.}
\end{figure}

The Rietveld refined XRD patterns for two samples of LaSrCoRuO$_{6}$ viz, LSCR13 and LSCR12 prepared at
1300$^\circ$C and 1200$^\circ$C respectively are presented in Fig. \ref{xrd06}. The stoichiometric double
perovskite, LaSrCoRuO$_{6}$ has a monoclinic structure with the B-site cations Co and Ru ordered in the NaCl
pattern in the space group P$2_1/n$. ND patterns recorded at 300K (Fig. \ref{nd}) show evidence for higher degree
of B-site order in LSCR13 as compared to LSCR12. The presence of the sharper (${1\over 2}$,${1\over 2}$,${1\over
2}$) super lattice reflection in the ND pattern in LSCR13 (see inset Fig \ref{nd}) indicates a higher degree of
ordering in LSCR13. Rietveld refinement of the XRD and ND patterns was carried out with $P2_1/n$ space group
wherein the La/Sr occupy the 4e site with fractional coordinates (0.0033, 0.0218, 0.25), Co is at 2c (0.5, 0,
0.5), Ru is at 2d (0.5, 0, 0) and the oxygen atoms occupy three sites, viz, (0.2886, 0.280, 0.0355); (0.2324,
0.774, 0.0264) and (-0.0662, 0.4938, 0.255) \cite{bos}. The scale factor, background parameters, cell parameters,
Co and Ru site occupancies along with instrumental broadening, totalling to 17 parameters were refined in that
order to obtain a good fit. The crystallographic parameters obtained from refinement of ND patterns along with
Curie-Weiss parameters calculated from magnetization measurements are summarized in Table \ref{riet}. Refinement
shows that there is only about 4\% disorder in the case of LSCR13 whereas in case of LSCR12 about 20\% of Co
occupies the Ru site (2d site) and vice versa thereby resulting in a larger disorder in the occupation of the
B-sites as compared to LSCR13. Therefore, we refer to LSCR13 as a ordered compound while LSCR12 is referred to as
disordered compound.

\begin{table}
\caption{\label{riet} Unit cell parameters, Co and Ru site occupancies obtained from Rietveld refinement and
Curie-Weiss parameters calculated from magnetization measurements at 1000 Oe for the two samples of
LaSrCo$_{1-x}$Ru$_{1+x}$O$_6$. Numbers in parentheses are uncertainty in the last digit.} \centering
\begin{tabular}{ccc}
\hline
~~~~~~Sample~~~~~~ &~~~~~~ LSCR13~~~~~~ &~~~~~~ LSCR12~~~~~~\\
a (\AA) & 5.5847(4) & 5.5891(3) \\
b (\AA) & 5.5592(6) & 5.5540(5) \\
c (\AA) & 7.8674(9) & 7.8787(5) \\
$\beta$ & 90.05(2) & 90.10(1)\\
Volume (\AA$^3$) & 244.25(4) & 244.57(3)  \\
Co ($1\over2$,0,$1\over2$) & 0.98(1) & 0.87(1)\\
Ru ($1\over2$,0,$1\over2$) & 0.02(1) & 0.13(1)\\
Ru ($1\over2$,0,0) & 0.98(1) & 0.87(1)\\
Co ($1\over2$,0,0) & 0.02(1) & 0.13(1)\\
$\mu_{eff}$ ($\mu_B$/fu) & 5.47(2) & 5.43(1) \\
$\Theta_{CW} (K)$ & -49(2) & -2.5(4) \\
 \hline
\end{tabular}
\end{table}

\begin{figure}
\centering
\includegraphics[scale=1]{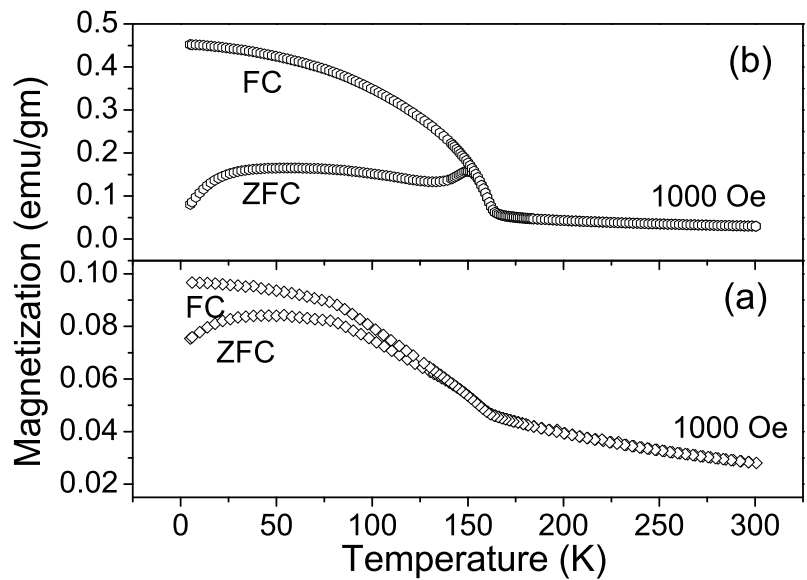}
\caption{\label{magn06} Magnetization as a function of temperature at applied fields of 1000 Oe in LSCR13 (a)
and LSCR12 (b).}
\end{figure}

Magnetization measurements performed at 1000 Oe during the ZFC and FC cycles for the two LaSrCoRuO$_{6}$ samples
are presented in Fig. \ref{magn06}. In case of LSCR13, both ZFC and FC cycles rise sharply below 160K and branch
off below 130K. While the ZFC curve culminates into a broad hump centered at about 55K, the FC curve approaches a
constant value below 77K. These curves do not reveal the nature of magnetic order in the compound. It may be
noted here that spin frozen ground state has been previously reported for this composition \cite{mam}. In yet
another study, antiferromagnetic order has also been have reported with T$_N$ = 87K \cite{bos}. In order to
confirm the nature of magnetic order in the present sample, ND pattern was recorded at 20K and is presented in
Fig. \ref{nd}. Weak extra reflections due to antiferromagnetic ordering are seen at the positions described by
propagation vector along the $k = {1\over2}~0~{1\over2}$ with respect to the crystallographic P$2_1$/n cell. This
magnetic arrangement is the same as that reported by Bos and Attfield \cite{bos}.

In the case of LSCR12 there is a wide difference in magnetization behaviour recorded during ZFC and FC cycles.
The ZFC magnetization with increasing temperature increases sharply culminating into a broad hump centred around
50K. It decreases slightly with further rise in temperature before increasing sharply resulting in a peak at
151K. The FC magnetization, on the other hand decreases continuously to 167K and settles into a low value giving
an impression of ferro to para transition. The differences in behaviour of magnetization during ZFC and FC cycle
indicates a complex magnetic ground state. ND pattern recorded at 20K also does not show any evidence of long
range magnetic order within our detectable limit. This could be implied to a magnetically frustrated ground state
due to presence of competing ferro and antiferromagnetic interactions.

\begin{figure}
\centering
\includegraphics[scale=1]{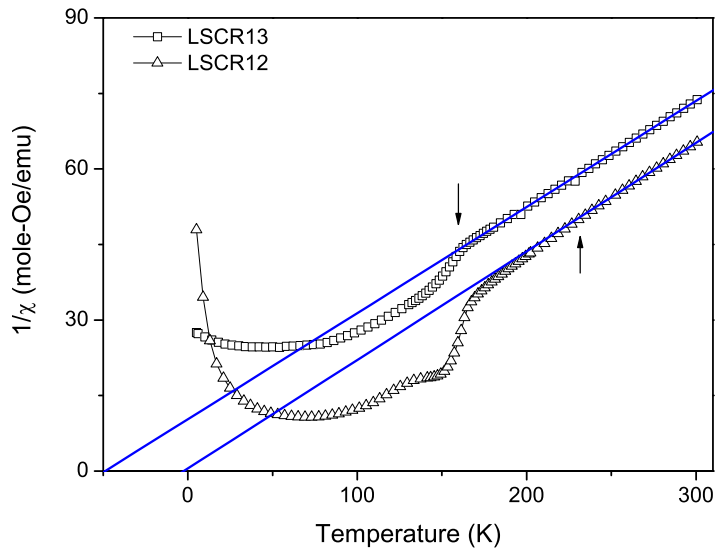}
\caption{\label{sus06} Inverse Magnetic susceptibility function of temperature calculated as $\chi$ = M/H at
applied fields of 1000 Oe in LSCR13 and LSCR12.}
\end{figure}

Plot of inverse of susceptibility (1/$\chi$ = H/M) in Fig. \ref{sus06}. For LSCR13 $1/\chi$ varies linearly with
temperature in the range 170K $<$ T $<$ 300K and Curie-Wiess fit to the data yields effective paramagnetic moment
$\mu_{eff}$ = 5.47 $\mu_B$/f.u. in good agreement with the calculated spin only moment of Co$^{2+}$ and Ru$^{5+}$
ions and the Curie-Weiss temperature, $\Theta_{CW}$ = -49K that is also in good agreement with the value reported
earlier \cite{mam}. The negative $\Theta_{CW}$ indicates presence of strong antiferromagnetic interactions. In
the case of LSCR12, the susceptibility although seems to be fairly linear down to 170K, deviates below the
Curie-Weiss behaviour at temperatures less than 220K. A linear fit in the temperature region 300K to 240K, to
inverse susceptibility of LSCR12 with Curie-Weiss equation  results in $\mu_{eff}$ = 5.43$\mu_B$/f.u. and
$\Theta_{CW}$ = -2.5K (see Fig. \ref{sus06}). The reduced value of $\Theta_{CW}$ and the deviation from
Curie-Weiss behaviour from higher temperature in LSCR12 points to presence of short range ferromagnetic
interactions in this compound. The presence of ferromagnetic interactions can be due to presence of small amount
of SrRuO$_3$ impurity which has an ordering temperature in the region of 140K - 160K. Although this agrees well
with the sharp rise in magnetization in LSCR12 below 170K, the deviation of susceptibility from Curie-Weiss
behaviour from about 220K hints at the presence of short range ferromagnetic interactions arising due to some
other reason than due to SrRuO$_3$ impurity alone. Further the absence of magnetic Bragg reflections in the
neutron diffraction pattern of LSCR12 due to antiferromagnetic order as in case of LSCR13 emphasize the presence
of short range ferromagnetic interactions within LaSrCoRuO$_6$ lattice. Presence of small amounts of SrRuO$_3$
impurity would not alter the magnetic ground state of parent LaSrCoRuO$_6$.

\begin{figure}
\centering
\includegraphics[scale=1]{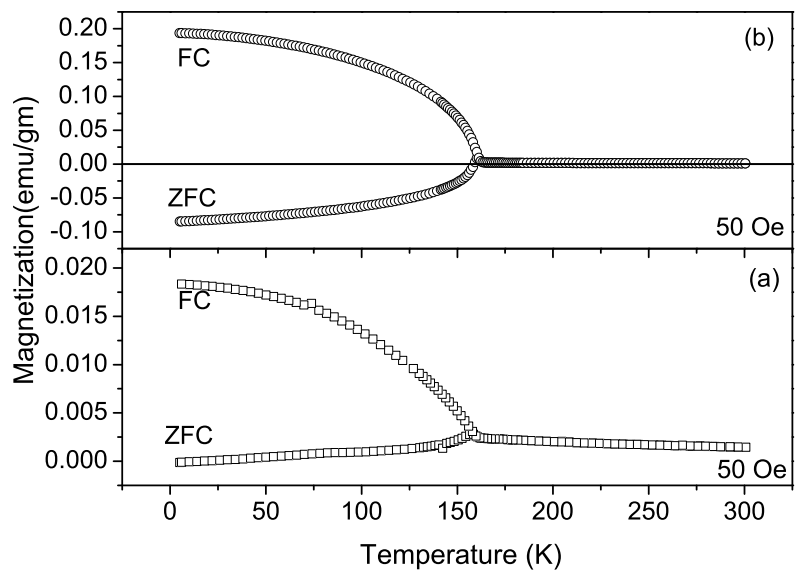}
\caption{\label{mag50} ZFC and FC magnetization curves at applied field of 50 Oe recorded for LSCR13 (a) and
LSCR12 (b).}
\end{figure}

In order to understand magnetic properties better, the low field (50 Oe) magnetization data were measured during
ZFC and FC cycles on the two samples of LaSrCoRuO$_6$ and are presented in Fig. \ref{mag50}. In case of LSCR12,
magnetization measured during the ZFC cycle is negative at the lowest temperature. It decreases in magnitude with
increasing temperature and crosses over to the positive side at 155 K, exhibits a peak at 160 K signifying a
transition from a magnetically ordered to paramagnetic state. During the FC cycle, magnetization behaviour is
similar except its value is positive throughout. Such a behaviour again cannot be understood to be due to
presence of SrRuO$_3$ impurity alone. It may be emphasized here that care has been taken to make sure that the
remanent field of SQUID magnetometer was less than $\pm$13 Oe during these low field measurements. In case of
LSCR13, although the magnetization exhibits significant deviation between ZFC and FC cycles below 160K but is
positive throughout.

\begin{figure}
\centering
\includegraphics[scale=0.5]{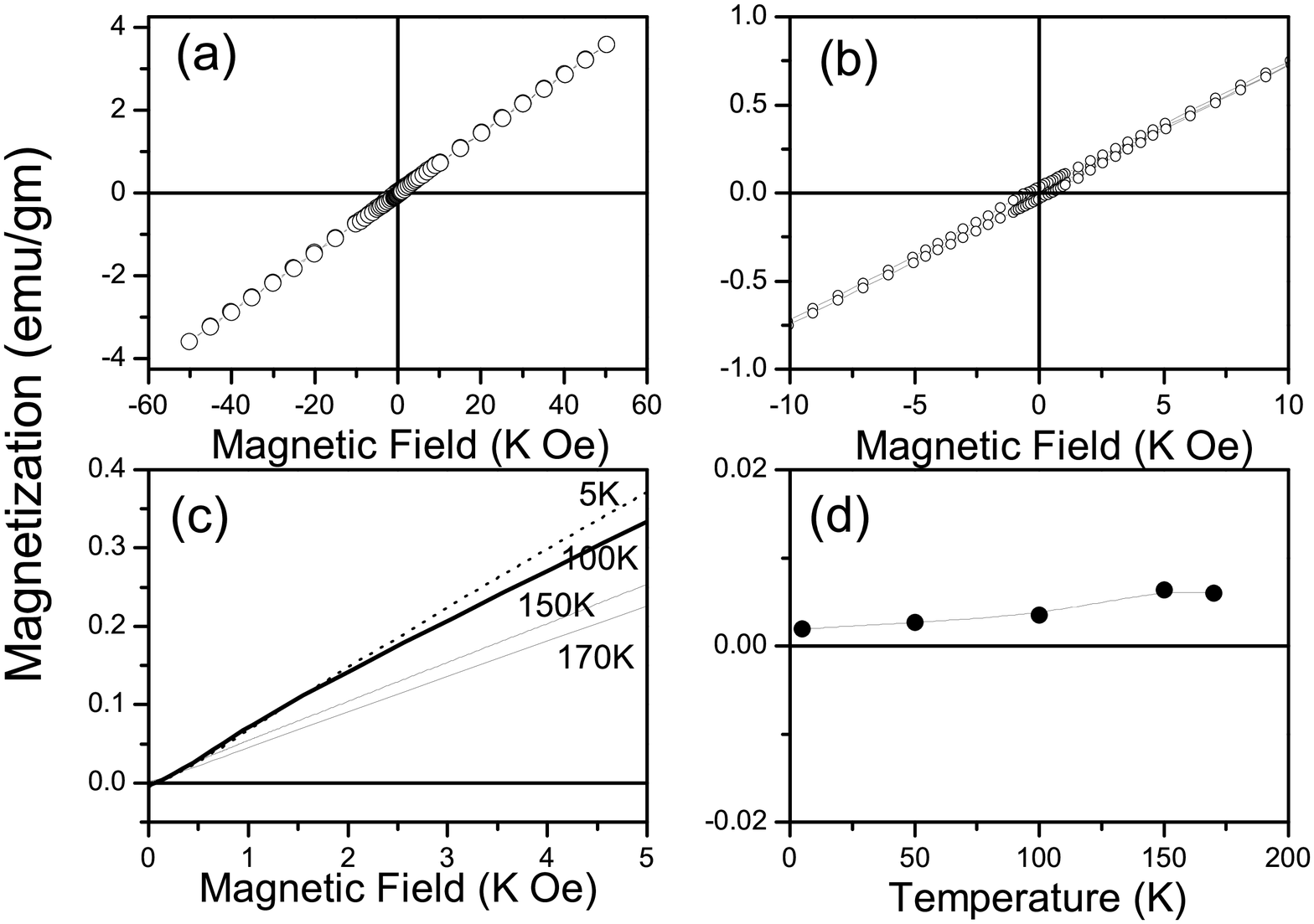}
\caption{\label{magnMH} Isothermal magnetization curves for LSCR13 recorded in the field interval of $\pm$5T at
5K (a); its magnified view ($\pm$10 KOe) (b); virgin magnetization curves at few representative temperatures (c)
and variation of magnetization values extracted from virgin curves at a field value of $sim$ 50 Oe.}
\end{figure}

\begin{figure}
\centering
\includegraphics[scale=0.5]{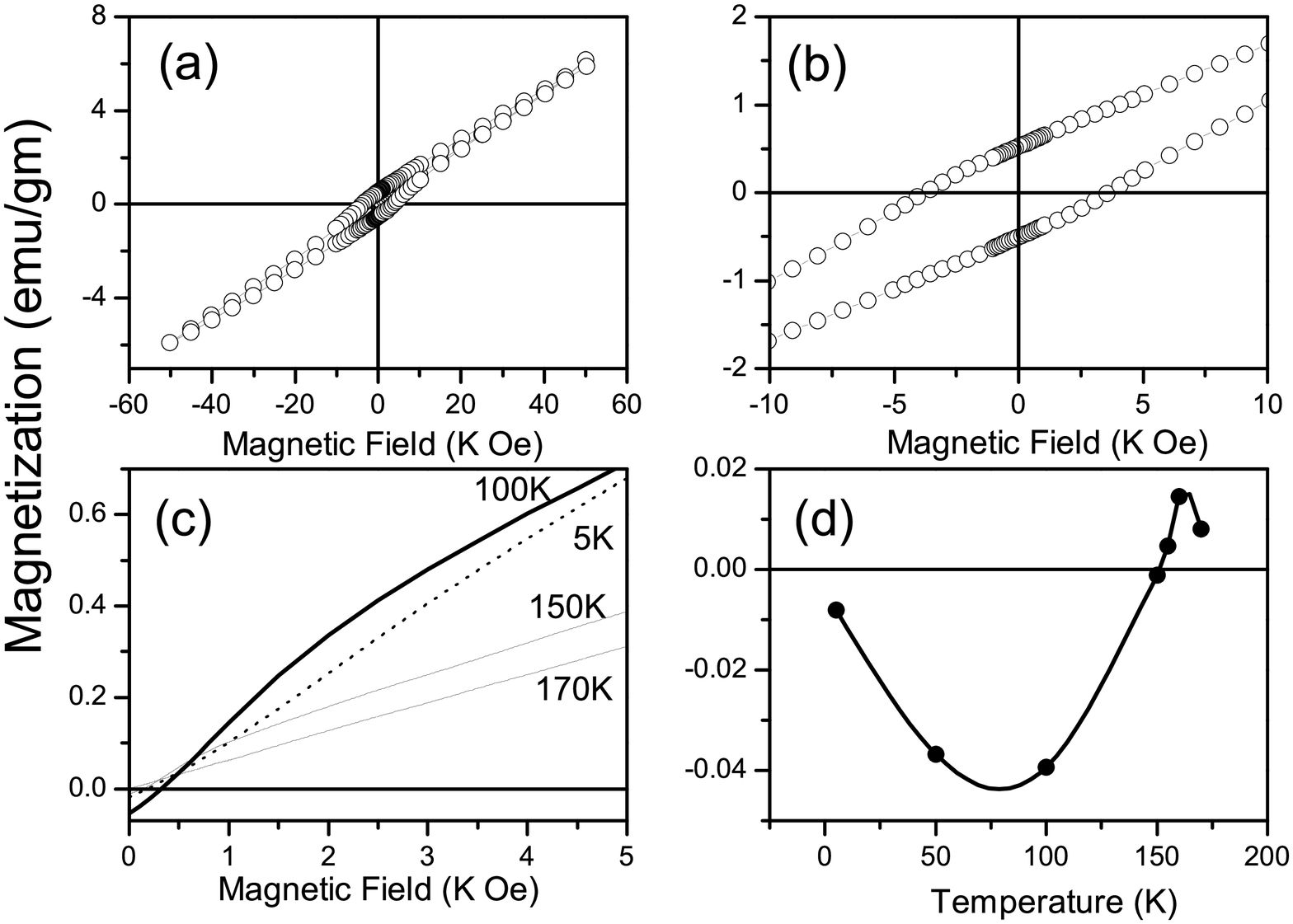}
\caption{\label{MH12} Isothermal magnetization curves for LSCR12 recorded in the field interval of $\pm$5T at 5K
(a); its magnified view ($\pm$10 KOe) (b); virgin magnetization curves at few representative temperatures (c) and
variation of magnetization values extracted from virgin curves at a field value of $\sim$ 50 Oe.}
\end{figure}

Isothermal magnetic response of the two samples has been studied at various temperature in the field range of
$\pm$ 50KOe. Fig. \ref{magnMH}(a) presents the isothermal magnetization curve for LSCR13 measured at 5K. The
amplified loop ($\pm$10 KOe) is presented in Fig. \ref{magnMH}(b). It can be seen that the magnetization exhibits
strong field dependency and almost no hysteresis which is typical of an antiferromagnet. On the other hand for
LSCR12 the isothermal magnetization studies performed at 5K (see Fig. \ref{MH12}(a)) exhibit a clear
ferromagnetic hysteresis loop riding on an antiferromagnetic (linear) background. Such a hysteresis loop is
typical for a compound with a ferromagnetic component along with antiferromagnetic interactions. The expanded
loop in Fig. \ref{MH12}(b) shows that the ferromagnetic component is quite strong and about 4 to 5 times larger
than that reported in Ref. \cite{bos}. Calculation of saturation moment by extrapolating the linear regions of
hysteresis loop yields a value of 0.64 emu/gm which is about 6\% of the value of Ru in SrRuO$_3$. This cannot be
ascribed to SrRuO$_3$ impurity alone as such a sizeable amount of SrRuO$_3$ would have been detected in
diffraction studies. Therefore the observed magnetic behaviour can only be ascribed to presence of competing
ferromagnetic and antiferromagnetic interactions resulting due to higher B-site disorder present in LSCR12. As
the low field magnetization measured in ZFC cycle is negative, it is worthwhile to see the behaviour of virgin
magnetization at different temperatures especially in the low field region. Fig. \ref{MH12}(c) exhibits the
virgin magnetization curves for LSCR12 and Fig. \ref{magnMH}(c) presents the same for LSCR13. While the
magnetization remains positive even very low fields for LSCR13, corresponding magnetization for LSCR12 is
negative. Further the shape of virgin curve at 100K makes it amply clear that ferromagnetism is more dominant
while the 5K curve is more linear corresponding to dominant antiferromagnetic interactions. This again excludes
the possibility of ferromagnetism arising due to SrRuO$_3$ impurity alone. Further, in case of LSCR12, the value
of magnetization at a low field ($\sim$ 50 Oe) extracted from virgin curves decreases below 150K and then shows a
upturn towards positive values below 80K. This is more clearly depicted in Fig. \ref{MH12}(d). This clearly
points to a presence of two magnetic sublattices which interact with each other leading to observed negative
magnetization. No such dependence is observed in case of LSCR13 (see Fig. \ref{magnMH}(d)).

\begin{figure}
\centering
\includegraphics[scale=0.5]{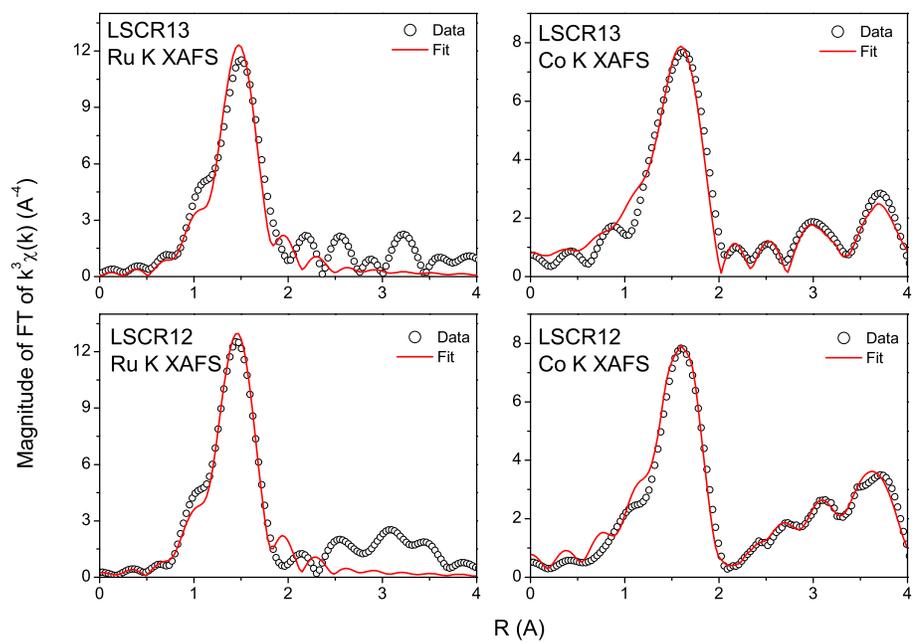}
\caption{\label{xafs-fig} k$^3$ weighted magnitude of Fourier transform of EXAFS data recorded at Co and Ru
K-edge in LSCR13 and LSCR12.}
\end{figure}

A disorder in site occupancy of Co and Ru sites can strengthen ferromagnetic interactions. Such a disorder can
result in Ru-O-Ru networks which will alter the Co and Ru octahedral networks, especially the Co-O-Ru bond angle.
In order to investigate the changes in the local structures around Co and Ru in between the two samples,
respective EXAFS data has been analyzed and the results are presented in Table \ref{xafs} and Fig.
\ref{xafs-fig}. It can be seen from the Table that in case of LSCR12 the Co-O and Ru-O bond lengths are lower and
the mean square radial displacements ($\sigma^2$) are higher as compared to those in LSCR13. Further, the values
of Co-Ru single scattering bond length and Co-O-Ru multiple scattering bond length indicate that the Co-O-Ru bond
angle increases in LSCR12 as compared to LSCR13. A straighter Co(Ru)-O-Ru(Co) bond angle implies a formation of
quasi-itinerant $\pi^*$ bands of Ru and/or ferromagnetic superexchange of high spin Co$^{2+}$-O-Ru$^{5+}$ type.
The higher value of $\sigma^2$ for Co-Ru bond distance are indicative of larger disorder in LSCR12. These local
structural changes can be understood to be due to B-site occupancy disorder in LSCR12 resulting in formation of
$\pi^*$ bands due to Ru-O-Ru linkages. These itinerant-electron $\pi^*$ bands interact ferromagnetically which
explains the sudden increase in magnetization below 150K. The ferromagnetic Ru sublattice so formed is coupled by
an antiferromagnetic exchange interaction to the Co-O-Ru antiferromagnetic sublattice. Below its ordering
temperature ($\sim$ 150K), the ferromagnetic Ru sublattice polarizes the paramagnetic Co moments in a direction
opposite to the applied field leading to magnetic compensation and negative magnetization. Once the Co moments
align antiferromagnetically below the antiferromagnetic ordering temperature of Co-O-Ru sublattice ($\sim$ 80K)
the magnetization increases towards a positive value as can be seen in Fig. \ref{MH12}(d).

\begin{table}
\caption{\label{xafs} Structural parameters like bond length (R\AA), bond angle and mean square radial
displacement ($\sigma^2$\AA$^2$) obtained from Co and Ru K edge EXAFS analysis. Numbers in parentheses are
uncertainty in the last digit.} \centering
\begin{tabular}{lcccc}
\hline
& \multicolumn{2}{c}{LSCR13}& \multicolumn{2}{c}{LSCR12}\\
Bond & R (\AA) & $\sigma^2$ (\AA$^2$) &  R (\AA) & $\sigma^2$ (\AA$^2$)\\
\hline
Co-O & 2.054(7) & 0.008(1) &  2.046(9) & 0.009(1) \\
Ru-O & 1.967(8) & 0.003(1) &  1.950(5) & 0.004(1)\\
Co-Ru & 3.97(6) & 0.002(1) &  3.97(1)& 0.005(1)\\
Co-Ru-O-Co & 4.00(7) & 0.002(1) &  4.00(1) & 0.004(1)\\
$\angle$ Co-O-Ru & 161.7(1)$^\circ$ &  & 166.4(1)$^\circ$ & \\
 \hline
\end{tabular}
\end{table}

LaSrCoRuO$_6$ is AA'BB'O$_6$ type double perovskite crystallizing in a monoclinic structure. This structure
allows for ordering of B-site cations in a NaCl fashion. This ordering is favoured due to the charge difference
($\Delta q \ge 3$) between Co and Ru. In perfectly ordered LaSrCoRuO$_6$, Co$^{2+}$ and Ru$^{5+}$ magnetic ions
couple antiferromagnetically leading to an antiferromagnetic ground state as can be seen from neutron diffraction
measurements. A disorder in Co and Ru site occupancy will result in Ru-O-Ru type linkages which are known to
align ferromagnetically. The presence of ferromagnetic interactions is clearly visible in LSCR12 which has a
larger B-site occupancy disorder in terms of increased values of magnetization as compared to those in LSCR13 and
hysteresis in M vs H loop. Due to such a disorder in occupancy of Co and Ru sites, octahedral distortions set in,
as the immediate neighbour of a Ru octahedra could be either a Ru octahedra or a Co octahedra. EXAFS results in
LSCR12 bear a testimony to this fact. In LSCR12, the Co-O and Ru-O bond lengths are shorter, the mean square
displacement is higher and Co-O-Ru bond is straighter. These changes are a result of Ru-O-Ru ferromagnetic
linkages which due to their presence alter Ru-O-Co antiferromagnetic interactions. The negative magnetization
seen in the low field ZFC magnetization is due to Ru-O-Ru ferromagnetic interactions which below $\sim$150K
polarize the paramagnetic Co spins in a the direction opposite to applied field giving rise to magnetic
compensation.

\section{Conclusions}
The disorder in occupation of Co and Ru sites in LaSrCoRuO$_6$ double perovskite results in Ru-O-Ru linkages
which are ferromagnetic. Due to such linkages, the magnetization of disordered compound increases in magnitude as
compared to that of ordered compound. At low applied fields the ferromagnetic spins polarize the paramagnetic Co
spins in a direction opposite to the direction of magnetic field resulting in observed negative magnetization.

\section*{Acknowledgements}
KRP and PRS would like to thank Department of Science and Technology (DST), Government of
India for financial support under the project No. SR/S2/CMP-42. PSRM acknowledges support from UGC-DAE Consortium
for Scientific Research, Mumbai Centre for financial support under CRS-M-126. KRP would also like to thank DST
for travel funding under Utilization of  International Synchrotron Radiation and Neutron Scattering facilities.


\newpage

\end{document}